\documentclass[12pt,aps,prb,preprint,showkeys,floatfix]{revtex4}   

\usepackage{amsmath}  
\usepackage{amsfonts}  
\usepackage{latexsym,amssymb}
\usepackage{graphicx}  
\usepackage{bm}

\draft
\begin{document}

\title{Complex behavior from a simple physical system:\\
A numerical investigation}

\author{Robert K. Murawski}
\email{rmurawsk@stevens.edu}
\affiliation{Stevens Institute of Technology\\ 
              Department of Physics and Engineering Physics\\ 
              Hoboken, N.J. 07030}

\date{\today}

\begin{abstract}
In this paper, the familiar problems of free-fall motion and simple harmonic 
motion (SHM) are combined. The novel composite system passes from 
regular to chaotic behavior for increasing values of energy $E$. 
This system is a suitable example to introduce undergraduates to some of 
the concepts of dynamical chaos theory.   
\end{abstract}

\keywords{chaos, spring, gravity, free-fall, Poincare map, 
Lyapunov exponent}

\maketitle

\section{Introduction}

For many centuries now, students of introductory physics have studied
Newton's Laws of Motion.  The two systems done ad nausea are a particle 
in free-fall and the mass-spring system (inclines being a special case
of the former).  Additionally, it is well know in the chaos community, over
the past few decades, that coupled oscillators can exhibit chaotic 
behavior i.e. motion that is very sensitive to initial conditions.\cite{tl75}
Examples of this kind include the driven damped oscillator,
the double pendulum, and the magnetic pendulum.  A few of these are
illustrated in Marion and Thornton's text on classical dynamics.\cite{mt95}

The purpose of this paper is to answer the question, ``What is the 
behavior of a mass-spring system which is subject to free-fall motion 
above a rigid floor without damping?''. This system is a variant of a 
class of problems know as impact oscillators.  One of the earliest was 
put forward by Fermi to explain the origin of cosmic rays.\cite{ef49} 
His model was an elastic ball which bounced off of a vertically 
oscillating floor.  The interested reader is referred to the 
bibliography for descriptions of similar systems.\cite{ws96,ld98} 

The free-fall spring subject to impact (FSI) constitutes a coupled oscillator 
problem and the resulting motion is chaotic.

\section{The Physical System: Coupled Oscillators}

Consider the motion of a point particle released from 
rest at an initial height $y$ subject to the acceleration due to 
gravity $g$. If the particle suffers a perfectly elastic collision 
with the floor (as is done in billiard theory) it will return to 
height $y$ after a time $T_{free-fall}$ given by 

\begin{equation}\label{E:1}
T_{free-fall} = \sqrt{\frac{8y}{g}}
\end{equation}

If the system is allowed to continually bounce, then it behaves like an
oscillator with frequency given by

\begin{equation}\label{E:2}
\nu_{free-fall} = \frac{1}{T_{free-fall}}
\end{equation}

Next, consider two equal masses at either end of a perfect spring with
an equilibrium position of $y_0$ and Hooke's constant $k$.  The 
motion of the difference coordinate i.e. $Y = y_2 - y_1$ will be SHM 
with frequency

\begin{equation}\label{E:3}
\nu_{SHM} = \frac{1}{2\pi}{\sqrt{\frac{2k}{m}}}
\end{equation}

\begin{figure}[ht]
\begin{center}
\includegraphics{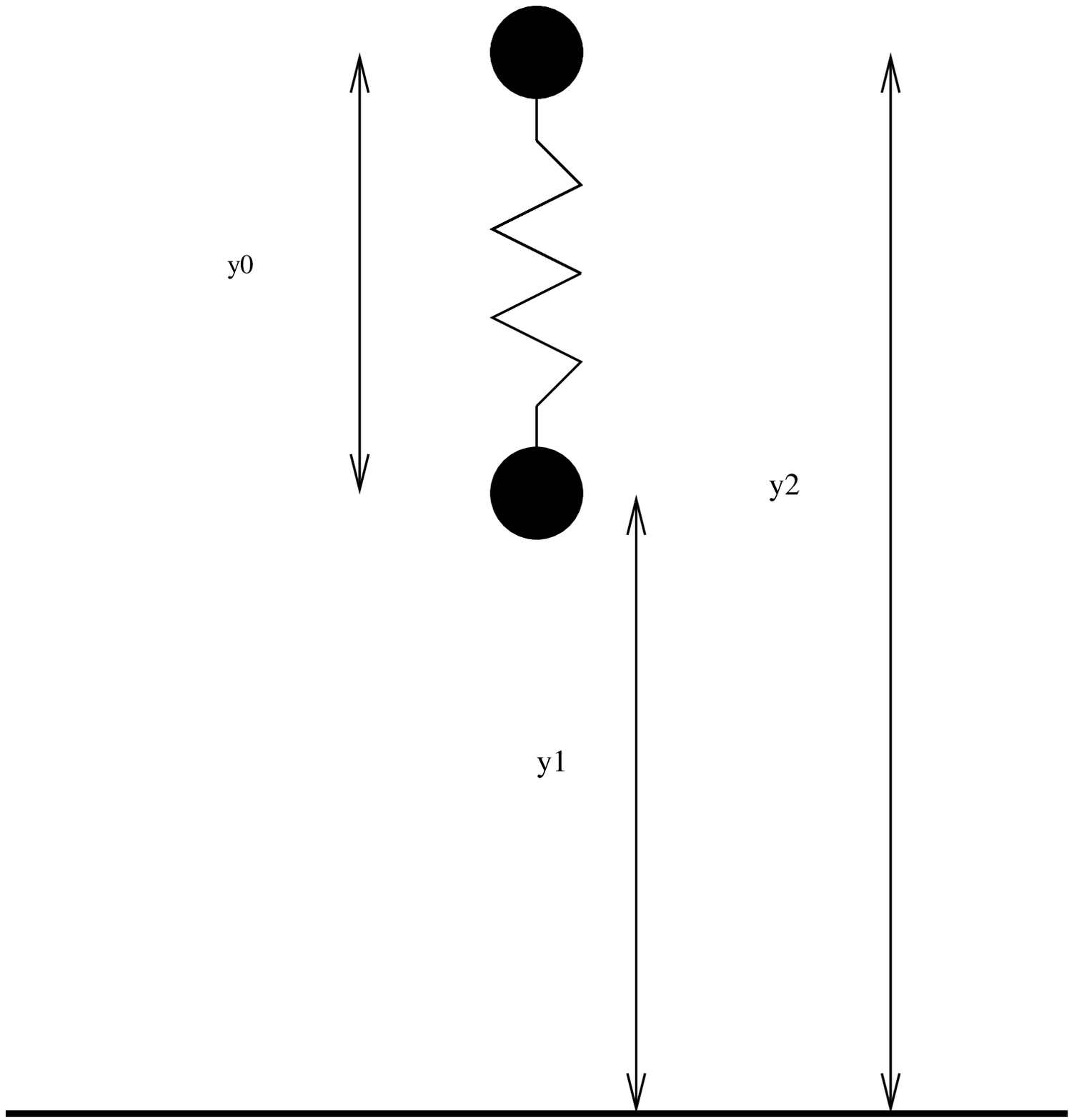}
\caption{\label{fig1}Two masses at either end of spring released vertically 
in a constant gravitational field over a perfectly reflecting surface}
\end{center}
\end{figure}

The motion of the two systems can be coupled by releasing the spring 
system vertically and having the lower mass reflect off of a 
perfectly elastic surface. The system is illustrated in figure 
\ref{fig1} and represents a coupled oscillator. Examples in nature 
of such a system would be a diatomic molecule of gas which suffers 
collisions with the floor of its container or a child on a pogo stick.

\section{Simulation: The Route from Physics to Computer}

In this section, the steps to get from physical model to a computer
simulation are outlined.  Dimensionless coordinates are also introduced.

\subsection{Equations of Motion:Model to Theory}

Newton's second law is used to write down the equations of motion. 
Taking the rigid floor to be at $y = 0$, ${m_1} = {m_2} = m$, and the 
rest length of the spring to be $y_0$ the accelerations of the respective
particles are  
  
\begin{subequations}
\label{E:Newton}
\begin{eqnarray}
\ddot{y}_1 &=& -g + {{\omega}_0}^2(y_2 - y_1 - y_0)\\
\ddot{y}_2 &=& -g - {{\omega}_0}^2(y_2 - y_1 - y_0)
\end{eqnarray}
\end{subequations}

The conventional dot notation has been used to represent the temporal
derivatives and ${{\omega}_0}^2 = k/m$ . To simplify the above 
equations \ref{E:Newton}, dimensionless coordinates are used

\begin{subequations}
\label{E:dim}      
\begin{eqnarray}
\frac{y_i}{y_0} \rightarrow {y_i}\\
t{\omega}_0 \rightarrow t\\
\beta = \frac{g}{{y_0}{{\omega}_0}^2} 
\end{eqnarray}
\end{subequations}
The dimensionless quantity $\beta$ contains all of the physical 
parameters of the problem. 

The two second order equations can be rewritten into a system of four first 
order equations by noting that velocity is the time derivative of position. 
In this way, and using system \ref{E:dim}, the dimensionless equations of 
motion for the FSI are
 
\begin{subequations}      
\label{E:eqm}
\begin{eqnarray}
\dot{y}_1 &=& v_1 \\
\dot{v}_1 &=& -\beta + (y_2 - y_1 - 1)\\
\dot{y}_2 &=& v_2 \\
\dot{v}_2 &=& -\beta - (y_2 - y_1 - 1)
\end{eqnarray}
\end{subequations}

The above system of equations ignores the interaction of the lower mass 
$y_1$ with the rigid floor. The lower mass receives a delta function impulse 
which switches its velocity $v_1 \rightarrow -v_1$ upon impact. Care must 
be taken on handling this impact and will be discussed in the numerics section.

\subsection{Energy Bounds}

There are energy bounds that must be taken into consideration for the 
free-fall spring subject to impact(FSI).First, the energy is written down in 
dimensionless form

\begin{equation}\label{E:Energy}
E = \frac{1}{2\beta}({v_1}^2 + {v_2}^2) + \frac{1}{2\beta}{(y_2 - y_1 - 1)}^2 + y_2 + y_1
\end{equation}

The terms are in the order of kinetic, spring potential, and gravitational 
potential energy.  The minimum of energy is chosen to be 
the system at rest with the lower mass on the floor at $y_1 = 0$ and the 
upper mass to be at $y_2 = 1 - \beta$ which corresponds to 

\begin{equation}\label{E:Emin}
E_{min} = 1 - \frac{\beta}{2}
\end{equation}

Since negative energy is unphysical in this problem, this relation 
requires that $\beta < 2$.  Moreover, $\beta$ must be chosen so that 
$y_2 > 0$ i.e. $\beta < 1$.  By choosing the spring to start off at 
its equilibrium length, the minimum of energy is $E_{min} = 1$. 

The upper bound on energy is found by requiring that the spring not be 
compressed to zero length i.e. $y_1 = y_2 = 0$. This is to ensure that 
particle two does not suffer a collision with the floor and that the 
two particles don't pass through each other. Substituting these initial
conditions into \ref{E:Energy} gives

\begin{equation}\label{E:Emax}
E_{max} = \frac{1}{2\beta}
\end{equation}

The energy and parameter requirements for the FSI are

\begin{subequations}
\label{E:Parameters}
\begin{eqnarray}
1 < E < \frac{1}{2\beta}\\
\beta < \frac{1}{2}
\end{eqnarray}
\end{subequations}

\subsection{Theory to Numerics}

The system was integrated in time with a fourth order Runge-Kutta 
method. The time step chosen was $h = 0.001$ which results in a local 
truncation error of approximately $\mathcal{O}(h^5) = 10^{-15}$. The 
energy of the system was monitored and noted to be conserved to eight 
decimal places in double precision arithmetic. For all the simulations
$\beta$ was chosen to be $\beta = 0.25$ making the energy range 
$1 < E < 2$. 

The subtle part of the simulation is the actual impact of the lower 
mass $y_1$ with the floor at $y = 0$. The time of impact can not be 
predetermined with a static discrete time step. As an example, let 
$y_1(t_i)$ equal the location of the lower mass right before impact at 
time $t_i$.  At the next time step $y_1(t_i +h)$ , the lower mass has 
moved through the floor and $y_1(t_i + h) < 0$ which is unphysical and 
a consequence of using discrete time steps. The situation can be 
remedied by switching the independent variable from time to $y_1$ and 
integrating the system back to the surface. Applying the chain rule 
from calculus, the system is transformed to

\begin{subequations}
\label{E:Transform}
\begin{eqnarray}
\frac{dy_1}{dy_1} = 1\\
\frac{dv_1}{dy_1} = \frac{\dot{v_1}}{v_1}\\
\frac{dy_2}{dy_1} = \frac{\dot{y_2}}{v_1}\\
\frac{dv_2}{dy_1} = \frac{\dot{v_2}}{v_1}
\end{eqnarray}
\end{subequations}

and integrated with h replaced by $h \rightarrow -y_1$. The system ,
in this way, is moved back by the amount of the overshoot. At this 
point, the velocity of the lower particle is switched $v_1 \rightarrow -v_1$ 
and the system is integrated forward again with the original time step.  
A similar example can be found in the references as well as a 
description of Runge-Kutta routines.\cite{km90}
  
\section{The Indicators of Chaos}

In this section, the system is investigated for the existence
of chaos and two popular chaos indicators are described.

\subsection{Sensitivity to Initial Conditions: Lyapunov Exponents}

Chaos has been described as a system which exhibits strong sensitivity
to initial conditions.  The famous ``butterfly effect'' of
Edward Lorenz is an example.\cite{el63}  In figure \ref{fig2a} two time series
of $y_2$ are plotted on the same graph.  Trajectory 1 is at an
energy of $E = 1.1$ and trajectory two at $E = 1.1 +\delta{E}$ where
$\delta{E} = 0.001$. The small separation in the two trajectories is 
slightly noticeable and the two paths follow each other closely. Also,
the trajectory looks periodic. When the energy is increased to $E = 1.75$
and $E = 1.75 + \delta{E}$ something very different happens.  The two
paths in figure \ref{fig2b} start out following each other but around $t = 10$ 
they start to diverge apart.  Also, the motion of neither looks periodic 
implying that higher harmonics have been introduced.  The proper way
to quantify this phenomena is with the Lyapunov characteristic
exponent.

If a system is chaotic, then two trajectories separated slightly 
in phase space will, on average, diverge away from each other with 
an exponential behavior given by

\begin{equation}\label{E:expo}
d(t) \approx {d_0}{e}^{\lambda t}
\end{equation} 

Where $d_0$ is the initial phase space separation at $t = 0$ and $\lambda$
is known as the Lyapunov characteristic exponent or the largest Lyapunov
exponent. Equation \ref{E:expo} should not be taken too literally but only 
represents an average behavior of dynamical systems.  The calculation of 
$\lambda$ is actually an iterative process.  

To start off, two trajectories initially separated in phase space by an 
amount $d_0$ are integrate forward for some time $\tau$.  The system is 
now separated in phase space by an amount $d_1$. At this point the two 
trajectories are renormalized so that the separation is again  $d_0$ and 
then integrated forward again for $\tau$.  Continuing in this manner, 
a set of N $d_i$ 's is used to calculate the Nth approximate of the 
Lyapunov exponent, given by

\begin{equation}\label{E:lya}
\lambda_N = \frac{1}{N\tau}\sum_{i=1}^N \ln{\frac{d_i}{d_0}}
\end{equation} 

The largest Lyapunov exponent is then given by taking the limit as 
$N \rightarrow \infty$ of equation \ref{E:lya}. The interested reader
should follow the references for a more detailed description.
\cite{mt95,gb76,gb80,sd90,je93}
  
In the presence of chaos, $\lambda$ tends toward some positive 
finite value. In the absence of chaos, $\lambda \rightarrow 0$. In 
figure \ref{fig3}, the results of $\lambda$ for the above two values of 
energy separated in phase space by $d_0 = 10^{-6}$ are shown. After $10^6$ 
iterations the results were $\lambda_{E=1.1} \approx 0.004$ and 
$\lambda_{E=1.75} \approx 0.187$.  The results imply that for the smaller 
energy the motion of the FSI is regular and for the larger energy 
the motion is chaotic. 

\subsection{Poincar\'{e} maps}

The system under consideration is four dimensional in phase space
given by the set $(y_1,y_2,v_1,v_2)$. Since energy is a conserved
quantity the degrees of freedom can be reduced to three.  A trajectory
through three dimensional phase space of the system would be impossible
to visualize and too complicated to follow. Information about the system
is still possible by cutting phase space with a plane and looking at
the crossing through this plane by trajectories at the same energy. 
The pattern left behind on the plane is known as a 
Poincar\'{e} map or a Poincar\'{e} surface of section.\cite{km90} 

In figure four, four maps for various values of energy are shown.
The surface of section plane was chosen by plotting points $(v_2,y_2)$
whenever $y_1$ made an impact with the floor. For the lowest value of
energy, $E = 1.1$, the motion is regular and periodic. As the energy
is increased to $E = 1.75$ the motion becomes more complex and the 
map becomes dense with points. This phase space filling behavior is
indicative of chaos.
 
\section{Conclusion}

The free-fall spring subject to impact is a simple model which exhibits
rich behavior.  The reader is invited to carry out other
investigations such as higher energies or setting $m_1 \neq m_2$.
The interested reader should see the references for non-technical\cite{jg87,fb89}
and technical introductions to chaos.\cite{eo93}

\begin{acknowledgments}
I would like to thank the Imperatore School of Science and Arts and
the Department of Physics and Engineering Physics at Stevens Institute
of Technology. In particular, I would like to thank Prof. George Schmidt
for helpful suggestions and Petra Sauer for proof reading this manuscript. 
\end{acknowledgments}

\newpage
\section*{The Physical System}

\newcounter{mycount}
\renewcommand{\thefigure}{\arabic{mycount}.\alph{figure}}
\setcounter{mycount}{2}
\setcounter{figure}{0}

\newpage
\begin{figure}[ht]
\begin{center}
\scalebox{0.75}{\includegraphics{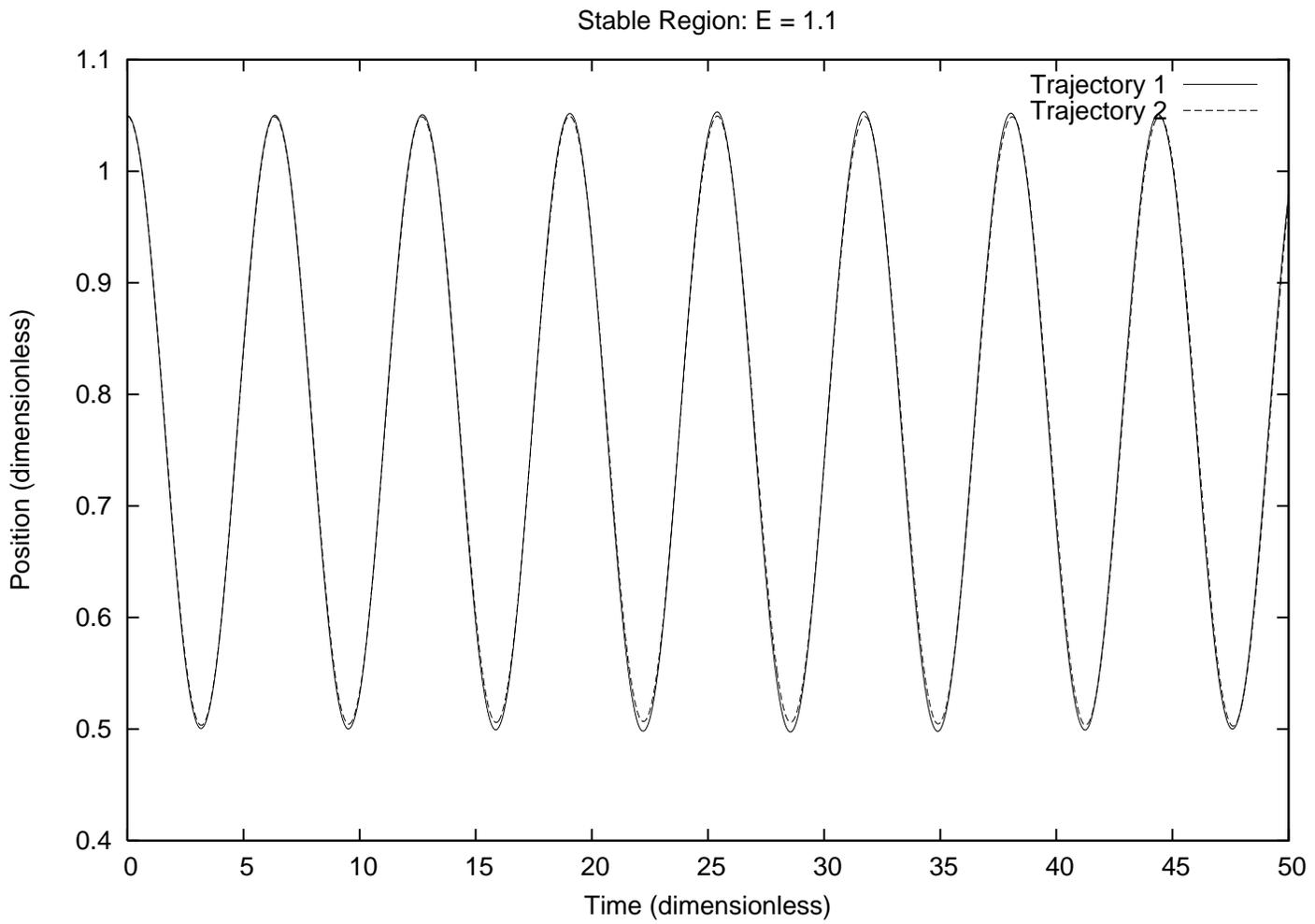}}
\caption{\label{fig2a}Trajectory 1 at E = 1.1
Trajectory 2 at E = 1.101}
\end{center}
\end{figure}

\newpage
\begin{figure}[ht]
\begin{center}
\scalebox{0.75}
{\includegraphics{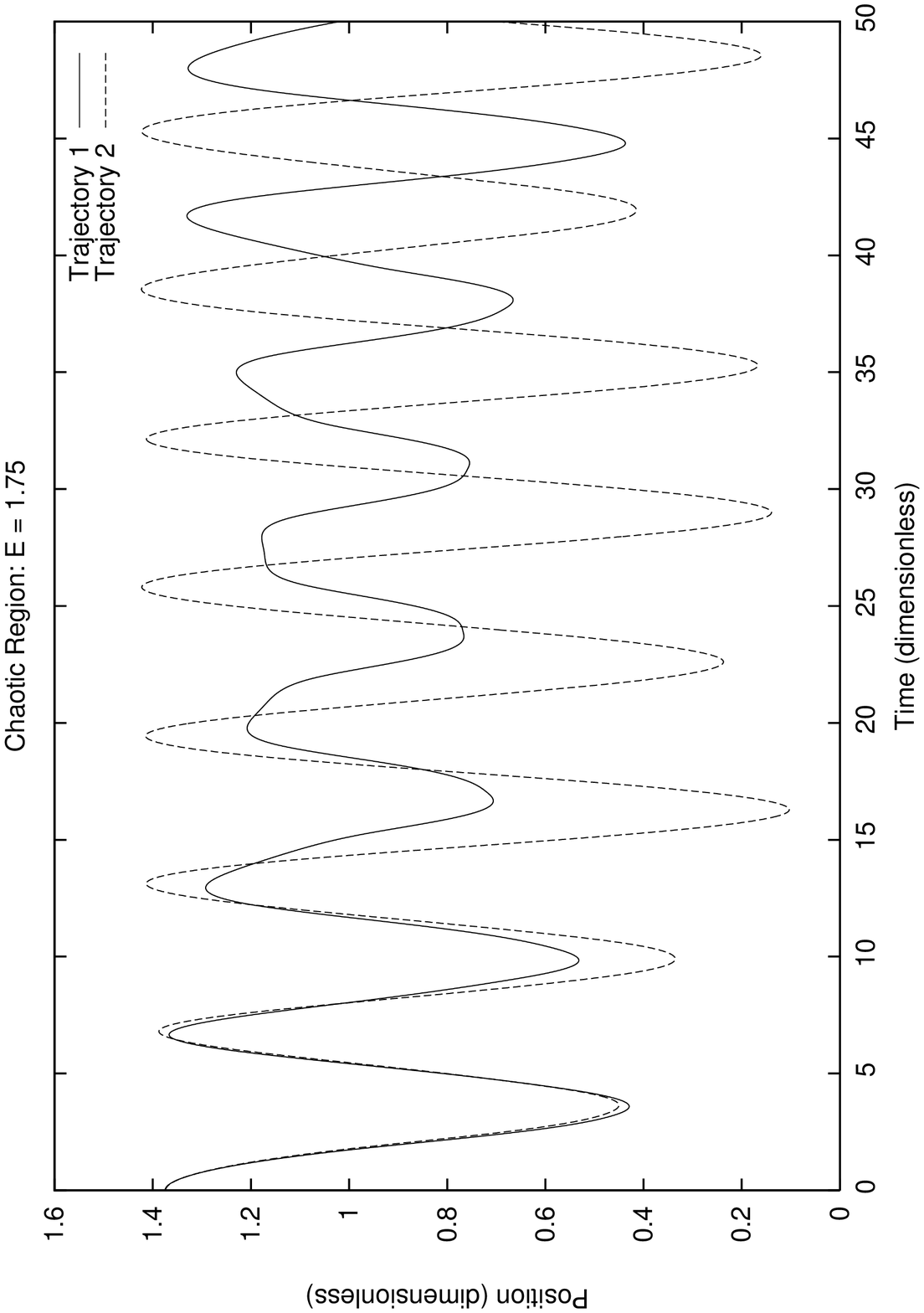}}
\caption{\label{fig2b}Trajectory 1 at E = 1.75
and Trajectory 2 at E = 1.751}
\end{center}
\end{figure}

\renewcommand{\thefigure}{\arabic{figure}}
\setcounter{figure}{2}

\newpage
\begin{figure}[ht]
\begin{center}
\scalebox{0.8}{\includegraphics{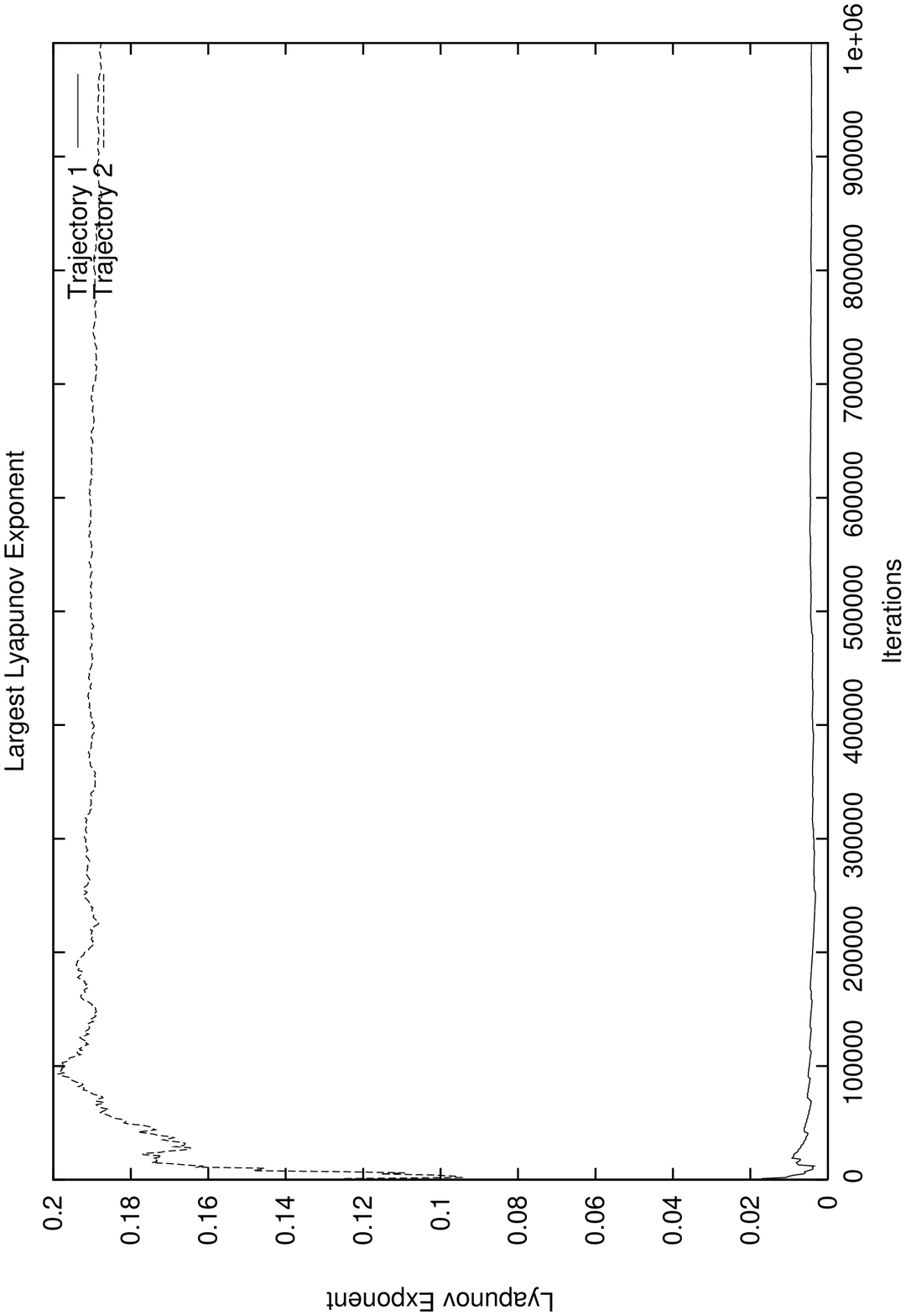}}
\caption{\label{fig3}$d_0 = 10^-6$ Traj.1 at E = 1.1
and Traj.2 at E = 1.75}
\end{center}
\end{figure}

\newcounter{mycount2}
\renewcommand{\thefigure}{\arabic{mycount2}.\alph{figure}}
\setcounter{mycount2}{4}
\setcounter{figure}{0}

\newpage
\begin{figure}[ht]
\begin{center}
\scalebox{0.8}{\includegraphics{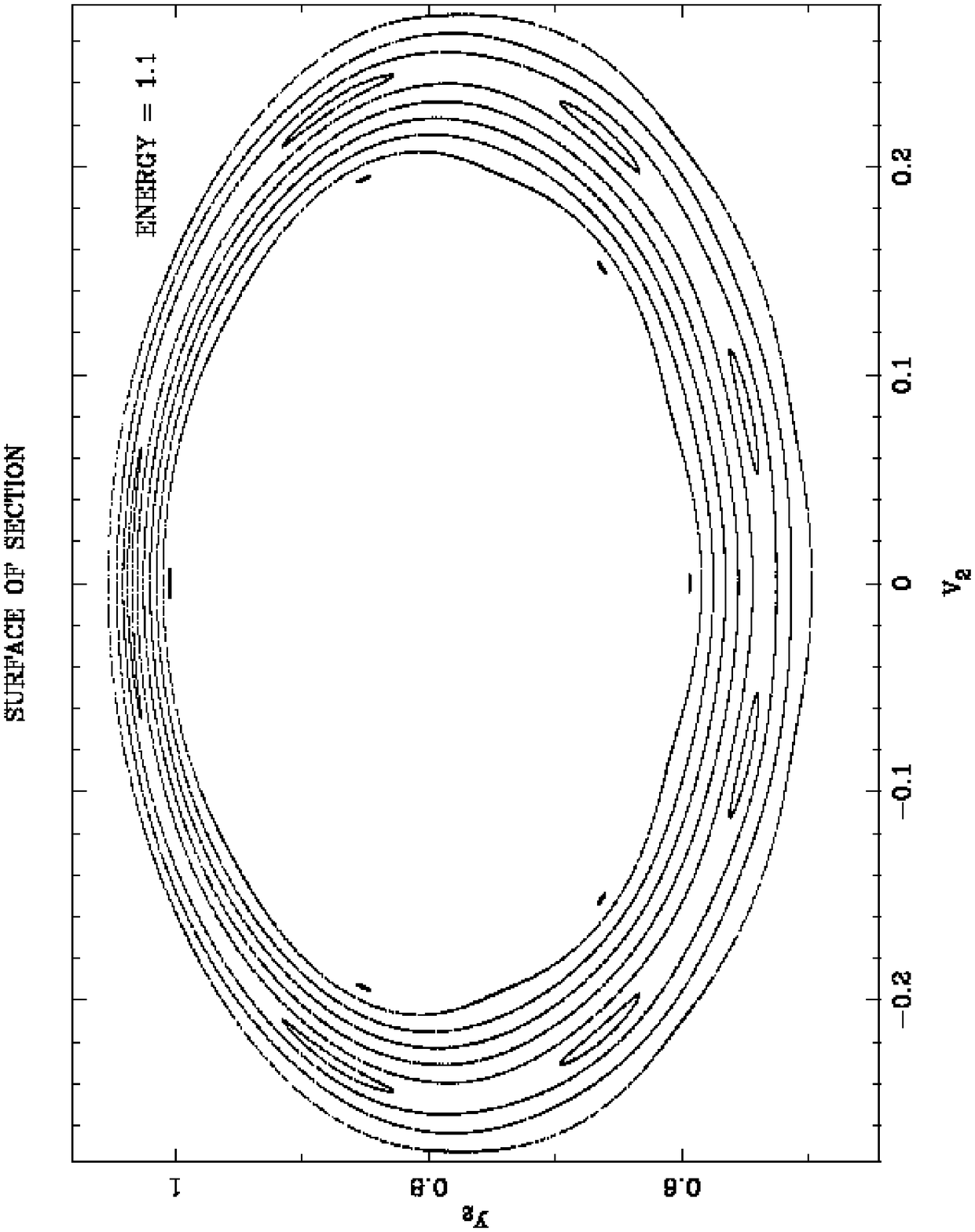}}
\caption{\label{fig4a} Energy = 1.1}
\end{center}
\end{figure}

\newpage
\begin{figure}[ht]
\begin{center}
\scalebox{0.8}{\includegraphics{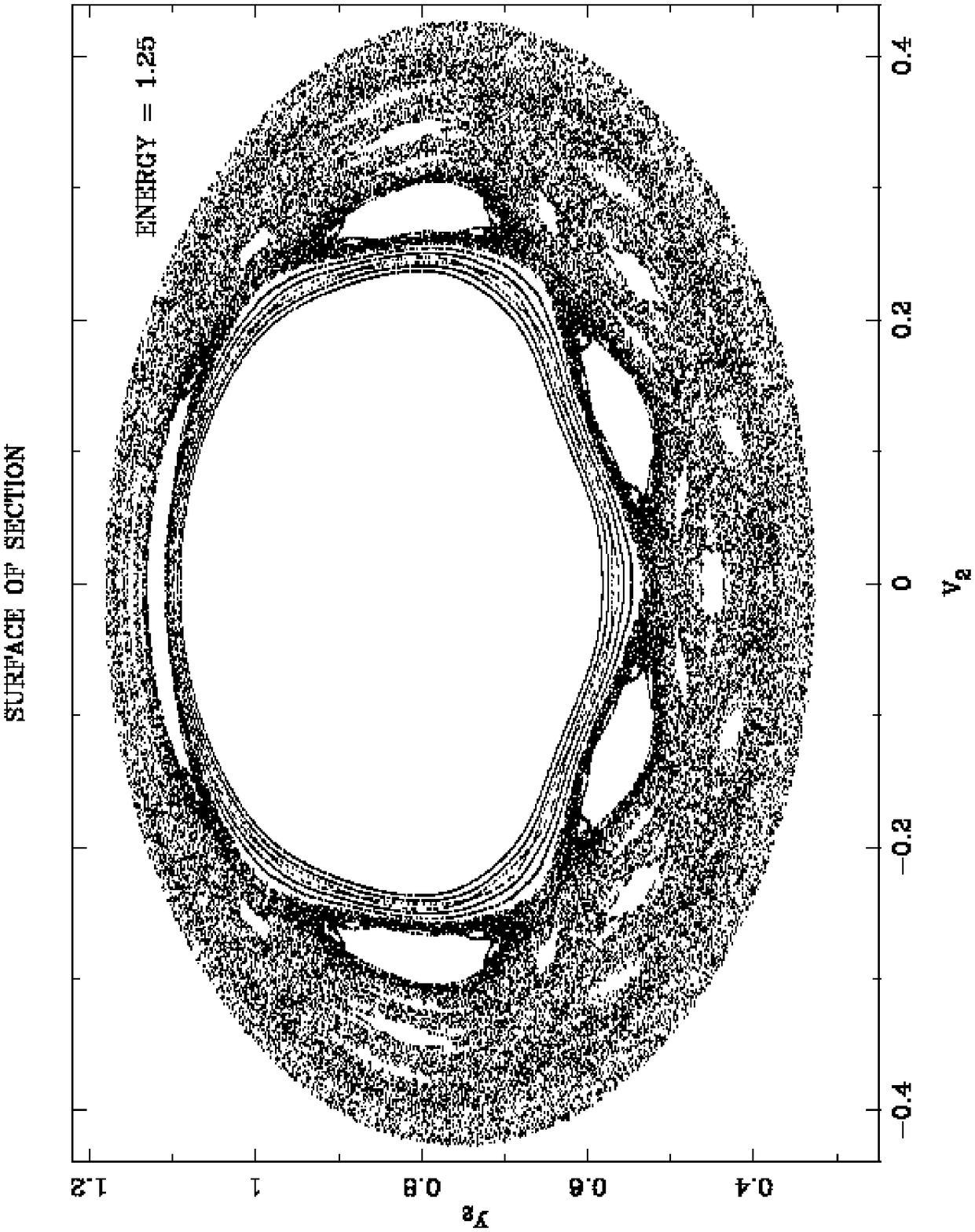}}
\caption{\label{fig4b} Energy = 1.25}
\end{center}
\end{figure}

\newpage
\begin{figure}[ht]
\begin{center}
\scalebox{0.8}{\includegraphics{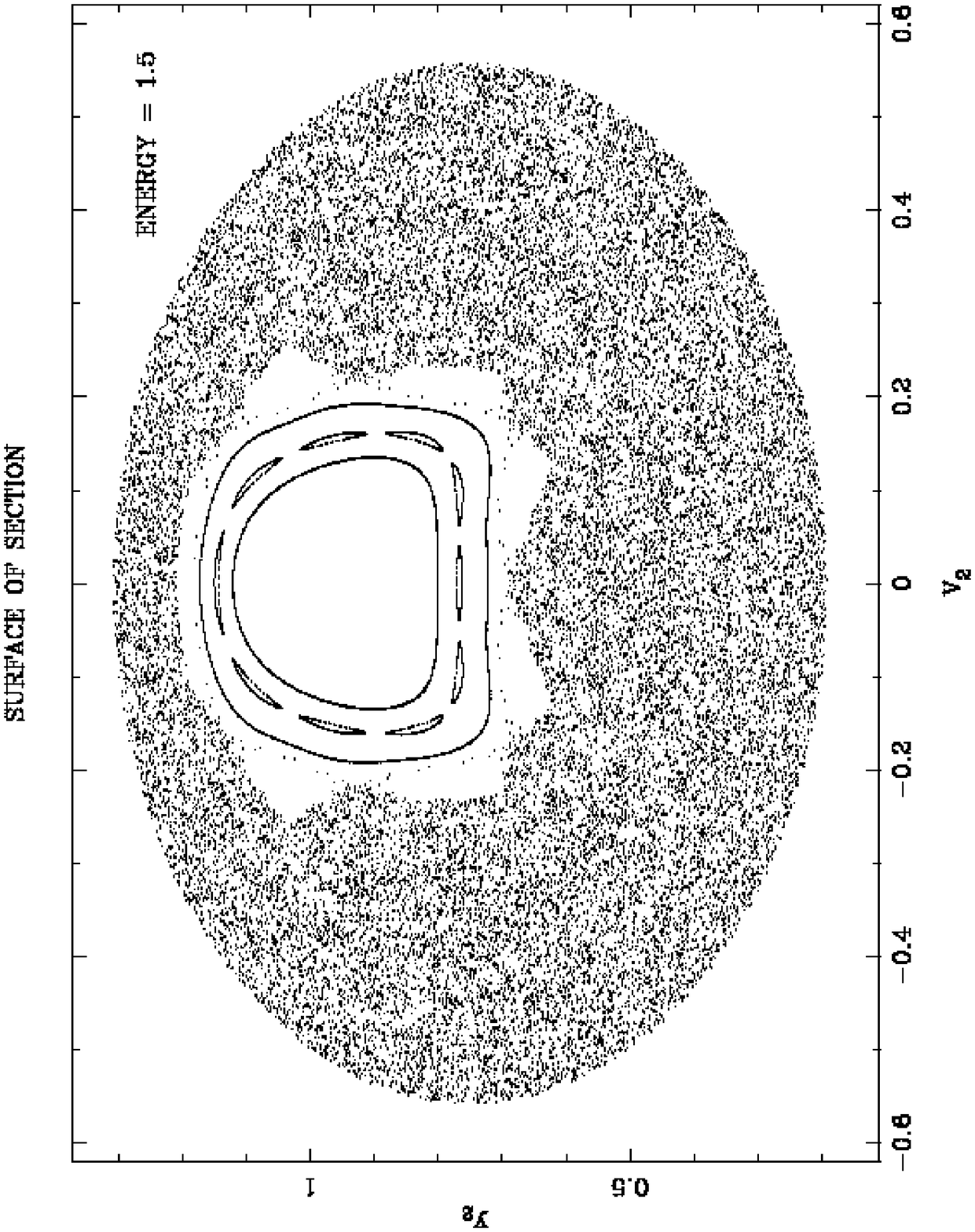}}
\caption{\label{fig4c} Energy = 1.5}
\end{center}
\end{figure}

\newpage
\begin{figure}[ht]
\begin{center}
\scalebox{0.8}{\includegraphics{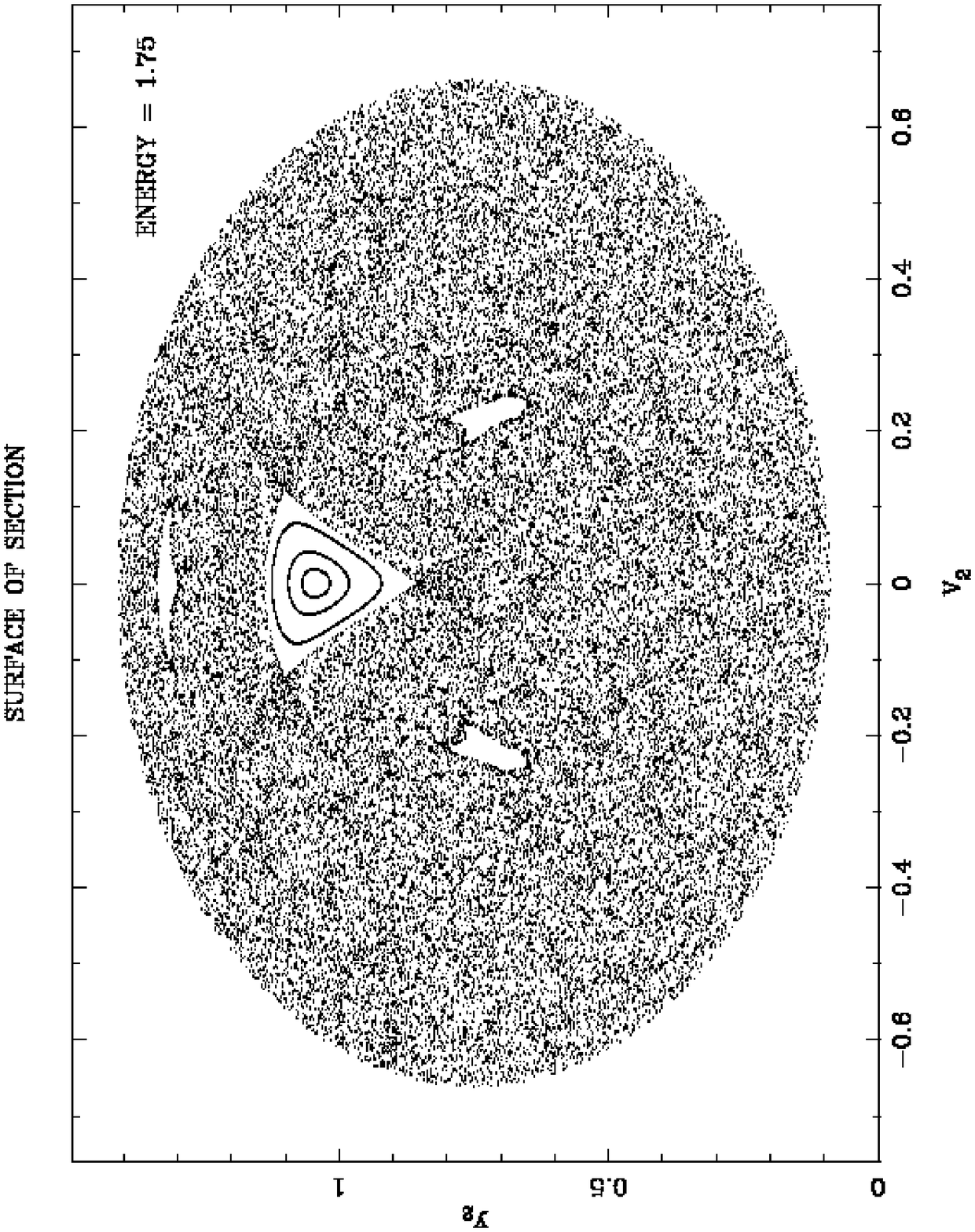}}
\caption{\label{fig4d} Energy = 1.75}
\end{center}
\end{figure}

\end{document}